\documentclass{article}
\usepackage[dvipdfmx]{graphicx}
\usepackage{arxiv}

\usepackage[utf8]{inputenc} 
\usepackage[T1]{fontenc}    
\usepackage{hyperref}       
\usepackage{url}            
\usepackage{amsfonts}       
\usepackage{nicefrac}       
\usepackage{cleveref}       
\usepackage{doi}

\usepackage{tikz}
\usetikzlibrary{intersections,calc,arrows.meta,math}

\usepackage{parskip}

\usepackage{booktabs}
\usepackage{threeparttable}
\usepackage{tabularx}
\usepackage{multirow}

\newcommand{\myPaperTitle}{Multimodal Deep Learning of Word-of-Mouth Text and Demographics to Predict Customer Rating: \\Handling Consumer Heterogeneity in Marketing
}
\newcommand{\myPaperShortTitle}{Multimodal Learning of Word-of-Mouth and Demographics}
\title{\myPaperTitle}

\date{}

\newif\ifuniqueAffiliation

\usepackage{authblk}

\newbox{\orcid}\sbox{\orcid}{\includegraphics[scale=0.06]{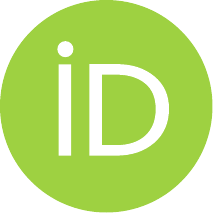}} 
\author[1,2]{%
	\href{https://orcid.org/0000-0002-4618-6272}{\usebox{\orcid}\hspace{1mm}
	Junichiro Niimi\thanks{\texttt{jniimi@meijo-u.ac.jp}}
	}}
\affil[1]{Meijo University}
\affil[2]{RIKEN AIP}

\renewcommand{\shorttitle}{\myPaperShortTitle}

\hypersetup{
pdftitle={Multimodal Deep Learning of Word-of-Mouth Text and Demographics to Predict Customer Rating: Handling Consumer Heterogeneity in Marketing},
pdfsubject={unsubjected},
pdfauthor={Junichiro ~Niimi},
pdfkeywords={Multimodal Learning, Deep Learning, Marketing},
}

\begin{document}
\maketitle

\begin{abstract}
In the marketing field, understanding consumer heterogeneity, which is the internal or psychological difference among consumers that cannot be captured by behavioral logs, has long been a critical challenge. However, a number of consumers today usually post their evaluation on the specific product on the online platform, which can be the valuable source of such unobservable differences among consumers. Several previous studies have shown the validity of the analysis on text modality, but on the other hand, such analyses may not necessarily demonstrate sufficient predictive accuracy for text alone, as they may not include information readily available from cross-sectional data, such as consumer profile data.  In addition, recent advances in machine learning techniques, such as large-scale language models (LLMs) and multimodal learning have made it possible to deal with the various kind of dataset simultaneously, including textual data and the traditional cross-sectional data, and the joint representations can be effectively obtained from multiple modalities. Therefore, this study constructs a product evaluation model that takes into account consumer heterogeneity by multimodal learning of online product reviews and consumer profile information. We also compare multiple models using different modalities or hyper-parameters to demonstrate the robustness of multimodal learning in marketing analysis.
\end{abstract}


\section{Introduction}
Deep learning is currently used in various fields. In the ﬁeld such as image recognition, it is already common to obtain features that affect the target variable by the feature extraction of deep learning \cite{handcraftedfeatures}, and it is considered one of the most powerful advantages of deep learning \cite{featureselection}. In the ﬁeld of customer relationship management (CRM) in marketing, it also has been widely employed in various ways of prediction and demonstrated the validity of utilizing deep learning, such as customer segmentation \cite{s2sLSTM-segmentation}, customer lifetime value (CLV) \cite{dnn-clv, dnn-clv2}, purchases (in the future or current session) \cite{dnn_purchase, dnn_purchase2, dnn_purchase_in_session}, churn \cite{dnn_churn1, dnn_churn2}, and other tasks \cite{lstm_directMarketing}. 

However, within the vast field of data science, a problematic factor, referred as consumer heterogeneity, has been identified, particularly in marketing. In marketing analysis, we combine a variety of behavioral log data; however, all the data recorded in the behavioral logs are the results of behaviors. The Differences in the psychological attributes among consumers that cause those behaviors cannot be obseerved by such logs. Many previous studies have highlighted the importance of considering heterogeneity \cite{hetero, hetero3}.

In terms of combining various data, multimodal learning \cite{multimodal1, multimodal2} has become widely popular in machine learning applications in general. It combines and learns different types of multiple data (i.e., modality), for example, audio and its corresponding text, in a state close to the original data. This method allows modeling that considers the relationships between modalities. However, in the marketing ﬁeld, analysis using such a variety of data is generally conducted by variablizing each data into a single dataset \cite{niimi_jsai}.

This paper is organized as follows. The previous studies are reviewed in Section 2, and the methodology (proposed model and used data) is shown in Section 3. An overview of the analysis is described in Section 4. The results of the analysis are presented in Section 5. Finally, we present the discussion and conclusions in Section 6.

\section{Related Work}
\subsection{Text Analysis in Marketing}
Natural Language Processing (NLP) techniques have been proposed for analyzing text modalities using deep learning. Among them, bidirectional encoder representations from transformers (BERT) \cite{bert} and their extensions \cite{distilbert, roberta} are some of the most popular methods because of their wide applicability, which is not limited to NLP tasks, such as machine translation and question answering. For pre-training, it has learned a contextual language representations using large text corpus; therefore, it can be further applied to various downstream tasks through ﬁne-tuning. Many models have already undergone pretraining with extensive data, enabling effective analysis with relatively small datasets. In recent years, an extension of BERT into multiple languages, known as multilingual BERT (mBERT), has emerged, and several models for the Japanese version, which is the focus of this study, have also been proposed (e.g., BERT models trained by Cyber Agent \cite{openCalm} and Tohoku NLP Group \cite{clTohoku}). Before the advent of BERT, vectorization methods such as word2vec \cite{word2vec} and doc2vec \cite{doc2vec} had been used to map sentences to feature vectors. Compared with techniques that produce a single- word embedding representation, the advantage of BERT is that it is context-wise, which means that it produces a representation based on other terms in the sentence \cite{bert, GooglePlay}.

For the actual use of these methods in marketing, especially regarding word-of-mouth texts, this study \cite{GooglePlay} predicts user review (that is, sentiment) for smartphone games using review texts collected from the Google Play Store, which compare several techniques to obtain word representations, including RNN \cite{RNN}, CNN \cite{CNN}, LSTM \cite{LSTM}, BERT, mBERT, DistilBERT \cite{distilbert}, and RoBERTa \cite{roberta}. In addition, a study \cite{bert_hotel} adopted BERT to predict customers’ ratings of hotels in seven criteria (e.g., overall ratings, value, and service) simultaneously, using online review texts for the recommender system. They indicated that BERT could predict a more accurate rating by considering the context of a review text. Another study regarding social media marketing \cite{bert_influencer} adopts BERT to capture the social media engagements and comments of influencers of eight categories on Instagram.  Thus, several studies have used BERT to map word-of-mouth documents to feature maps to predict customer evaluations.

In addition, while BERT has two different scale: Base (approx. 110 million parameters) and Large (approx. 340 million parameters), the extent to which such scales of BERT affect the result of the analysis for relatively small text data is yet to be clariﬁed, particularly in non-English models\footnote{The results are comparable between the model scales since both BERT models of Base and Large are designed to handle the same number of input tokens.}.

\subsection{Multimodal Learning in Marketing}
Multimodal learning involves learning representations from multiple modalities such as images, videos, audio, and text. A combination of these enables the construction of a robust learner based on the relationships among modalities that  cannot be obtained by learning a single modality \cite{multimodal2}. There are a variety of applications, including the integration of information from multiple sources and interconversion between modalities, applied to a wide range of academic ﬁelds such as medicine, human-computer interaction, biometrics, and remote sensing \cite{multimodal-survey}. In contrast to this rise in multimodal learning, research on multimodal learning in the marketing field is relatively scarce. One possible reason is that marketing data analysis involves a mixture of various data in different formats, such as server logs, ID-POS, GPS, survey responses, and customer information. Although analyses combining various datasets are widely practiced, they are typically conducted by variablizing multiple data and merging them into a single set for analysis
\footnote{The information loss occurring in the process of converting behavioral log into cross-sectional data may lead to a decrease in the accuracy of marketing analysis using deep learning. In other words, the manually variablized features cannot be sufficient statistics for the task on their own.}. 

However, there are a few notable instances, such as studies that construct a multimodal deep learning model to predict consumer loyalty with the source–target attention mechanism \cite{alaraj}, which datasets with different dimensionality are input simultaneously; however, by using bidirectional LSTM, the time-series data is converted to two-dimensional and uniﬁed to a single representation with cross-sectional data by feature fusion\footnote{Notably, in both studies, data fusion is conducted twice in one network structure: source-target attention and feature fusion.}. 

\subsection{Consumer Heterogeneity in Marketing}
Both the studies mentioned above highlight that it is possible to enhance discriminative power by considering demographic variables as a context affecting actual behavior. In the CRM context, the problem of consumer heterogeneity has long been highlighted, where even consumers who perform same behavior have unobservable differences \cite{hetero}, such as demographic and psychographic differences. In general, this kind of difference is unobservable in behavioral logs such as ID–POS data. Several studies have attempted to capture such differences using statistical modeling, such as structural equation modeling (SEM) and Bayesian mixture models \cite{hetero1, hetero2}. When it comes to the evaluation of the results addressed in this study, even among customers who have the same rating (number of stars), the reasons for the rating must differ; however, this difference cannot be observed in the cross-sectional data. In recent years, with the availability of a variety of large-scale data, document data such as word-of-mouth on online platforms have been used in the aforementioned studies as an important source for understanding consumer preferences \cite{review_for_heterogeneity}. 

Thus, several studies have utilized BERT and text data to make predictions that account for consumer heterogeneity. In general, the review text has more or less amount of reason why they evaluate the venue in such rating. In other words, the data, although it is behavioral log, can be the important source of understanding consumer heterogeneity. Therefore, this study utilize product review as a means to capture consumer heterogeneity to predict with better performance.

On the other hand, relying sorely on the feature extraction of machine learning is not advisable because domain knowledge is not incorporated into the analysis. In fact, several studies have shown that in multimodal learning, the combination of extracted features and handcrafted variables achieves the best prediction accuracy \cite{handcraftedfeatures, niimi_jsai}. Therefore, this study constructs a multimodal deep learning model that combines the review text with handcrafted user profile variables to achieve a robust and precise model.

\section{Proposed Model}
Based on previous studies of multimodal learning using time series and cross-sectional data [17], we present a conceptual model of multimodal deep learning that integrates three smaller neural network components, referred to as subnetworks or subnets. In this study, they are called text-speciﬁc subnetworks (X1-subnet), cross-sectional-data-speciﬁc subnetworks (X2-subnet), and output subnetworks (output-subnet). The construction of each subnet is described in the following subsections.

\begin{figure}[t]
     \centering
      \includegraphics{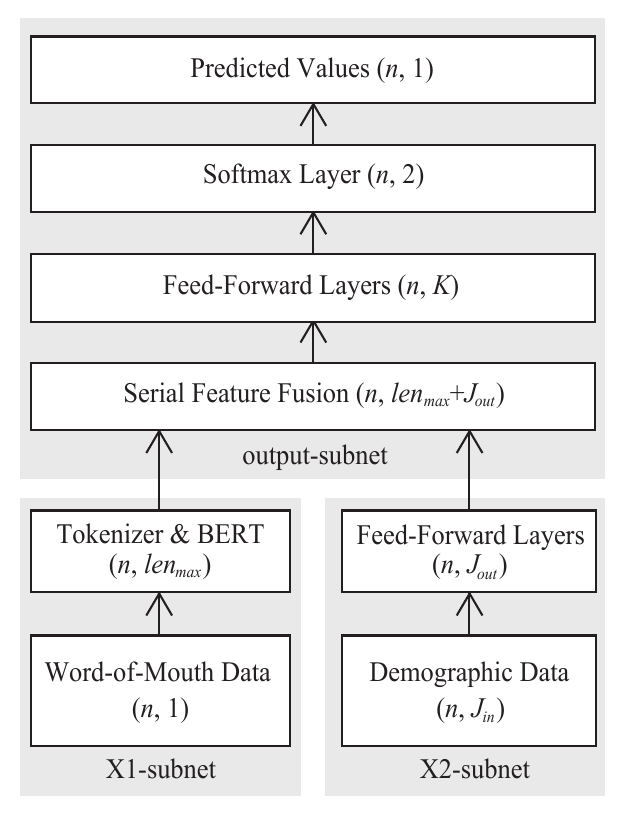}
      \caption{Model Architecture}
      \label{fig:architecture}
\end{figure}

\subsection{Text-Specific Subnetworks (X1-subnet)}
First, we describe the structure of the X1-subnet, which specializes in processing text data. The purpose of this subnet is to map word-of-mouth texts (whose lengths differ among users) onto a two-dimensional single-feature map. The actual component included two layers: a tokenization layer and a BERT layer. In this study, BERT was used to obtain embedded representations of text data; therefore, the X1-subnet (tokenizer and BERT layers) is freezed throughout the actual training process, which means that the parameters of the subnet are held fixed in the pre-trained state.

The text data contains the review of $N$ users and is processed with tokenizer into ﬁxed-length of ($N,~len_{max}$) tensor ($len_{max}$ represents the maximum number of tokens of the text among $N$ sets of data). After processing in the BERT layer with batch size $n$, we obtained the feature map as the pooler output of BERT as the ($n, len_{max}$) tensor.\footnote{The maximum length of the text $len_{max} \le 512$ since it cannot be exceeded the maximum number of tokens that BERT can deal with.}

\subsection{Cross-Sectional-Data-Specific Subnetworks (X2-subnet)}
Next, we describe the structure of the X2-subnet handling two-dimensional cross-sectional data. This subnet consists of a typical deep neural network with feed-forward layers (FFLs). An input layer for the X2-subnet receives a 2-dimensional tensor ($n, ~J_{in}$), with batch size $n$ and $J_{in}$ variables in the cross-sectional data. After processing, ($n, ~J_{out}$) feature map is obtained as the output. 

\subsection{Output Subnetworks (output-subnet)}
Textual and cross-sectional data, processed in parallel in each dedicated subnet are sent to the output subnet. In this subnet, both the feature maps, stemming from X1 and X2, are uniﬁed as a joint representation by feature fusion layer\footnote{However, a more rigorous call of feature fusion in this study should be "intermediate fusion" \cite{multimodal-survey}.}, and the obtained feature map would be ($n, ~len_{max}+J_{out}$) tensor. After the feature fusion, the joint representation is processed through one or more FFLs, and finally classification is conducted with Softmax layer. All the FFLs throughout the model employed a hyperbolic tangent function ($tanh$) for layer activation.

\section{Analysis}
\subsection{Dataset}
To conduct this analysis, we randomly selected one product from the women’s cosmetics market. The target product had to be well-recognized in Japan and already out of production. We collected data from 1040 participants online. The actual data contains three kinds of modalities: rating for the product (7-point Likert scale), word-of-mouth texts for the product, and demographic information \footnote{Since the cosmetic we focus on is a product intended for use by females, the survey was limited to females who purchased the products themselves.}. Finally, the sample size is 1532 (i.e., $N=1532$).

Next, we process the survey text in the general way of preprocessing in NLP. The text contained line breaks, pictographs, emoticons, and other characters that were not appropriate for analysis. Therefore, in the preprocessing stage, these elements were replaced with periods only when they were placed at the end of the sentence; otherwise, the elements were removed. Subsequently, all successive punctuation periods are merged into a single period. Consequently, the maximum length of the text data was set as $len_{max} = 200$.

\subsection{Model Evaluation}
This study adopts a simple binary classiﬁcation for the task, similar to previous studies \cite{GooglePlay}. User ratings were dichotomized into two classes based on the rating scale. Six and seven stars were classiﬁed as $Loyalty=1$ (loyalty is high) and $Loyalty=0$ (loyalty is not high), respectively. The obtained dataset was divided into a training set (75\%) and a test set (25\%). The model performance was evaluated using both training and test accuracies and the number of epochs to converge.

This study aims to validate several key points to compare the models. First, to validate the usefulness of multimodal learning in marketing, we construct three basic models according to their modalities: X1-modal, X2-modal, and multimodal. This comparison allowed for the veriﬁcation of changes in prediction accuracy by combining multiple modalities. Especially in marketing analysis, review texts can be a valuable source for comprehending consumer heterogeneity in user ratings.

Next, we examined the change in prediction accuracy using multiple pre-trained models within BERT. Regarding the Japanese language model, several models of different scales (that is, numbers of parameters), different training datasets, and tokenization methods (in particular, with regard to Japanese models, some are trained on a word-by-word basis, whereas others are trained on a character-by-character basis) have already been proposed, and the extent to which the prediction accuracy differs depending on the use of these models is yet to be clarified. Nowadays, we can easily switch the pre-trained model in BERT by changing one line of the code, which makes it easy to compare the accuracies of different pre-trained models. As it has already been shown in the literatures \cite{GooglePlay, bert_hotel, bert_influencer} that models using BERT-like architectures achieve higher accuracy compared to those using Collaborative Filtering, LSTM, CNN, and other benchmark models, this study sticks to comparing among multiple models using BERT.

The model settings are shown in Table \ref{tab:explore}. In the case where several candidates are shown in the setting, we utilize a grid search to explore the settings that maximize the test accuracy. For example, we compared the prediction accuracy among the four pre-trained models in BERT (bert base/large in word/char) and among three optimizers (Adaptive Moment Estimation, Adam \cite{adam}, Adamax \cite{adam}, and Nesterov-accelerated Adaptive Moment Estimation, Nadam \cite{nadam}).

The training process was conducted with a maximum of 200 epochs and 64 batch sizes. Early Stopping \cite{earlyStopping} was employed with a patience of 50 epochs, which terminates the training if no improvement in the accuracy of the validation data was observed within 50 epochs.

\begin{table*}[t] 
      \begin{center}
      \caption{Model Settings}
      \label{tab:explore}
        \scalebox{1}{
        \begin{threeparttable}
\begin{tabularx}{\linewidth}{llll}
\toprule
~~~
&\multicolumn{2}{c}{\bf Parameters}        &   \multicolumn{1}{c}{\bf Candidates} \\
\midrule
&\multicolumn{3}{l}{\bf Model Parameters} \\

&~~~~& Number of Epochs     &     200 (with Early-Stopping in 50 Epochs)~~~~~ \\
&& Batchsize ($n$)      &         64 \\
&& Optimizer                &  \{Adam, Adamax, Nadam\} \\
&& Loss Function         &  Binary Cross-entropy \\
\midrule
&\multicolumn{3}{l}{{\bf X1-subnet}} \\
&& Structure & mBERT (Japanese) \\ 
&& Model      & \{bert-base-japanese-v3,\\
&&                 & ~~bert-base-japanese-char-v3,~~~~~~~~~~ \\ 
&&                 & ~~bert-large-japanese-v2,  \\ 
&&                 & ~~bert-large-japanese-char-v2\} \\ 
\midrule
&\multicolumn{3}{l}{{\bf X2-subnet}} \\
&& Number of Hidden Layers  &  2 \\
&& Number of Neurons in the Layer~~~~~~~~~~~~~~~      &  10 \\
&& Activation Function &        tanh \\
\midrule
&\multicolumn{3}{l}{{\bf output-subnet}} \\
&& Number of Hidden Layers   &             2 \\
&& Number of Neurons in the Layer      &    10 \\
&& Activation Function &        tanh \\
\bottomrule
\end{tabularx}
\begin{tablenotes}[para,flushleft,online,normal] 
{\it Note.} tanh stands for hyperbolic tangent function.
\end{tablenotes}
\end{threeparttable}
        }
        \end{center}
\end{table*} 

\section{Results}
The best models for each modality are listed in Table \ref{tab:results1}. First, in both the training and test results, the prediction accuracy improved the most with multimodal learning. Although these results do not allow us to evaluate whether multimodal learning immediately improves the prediction accuracy, multimodal learning with the bert-base-japanese-v3 model shows the highest prediction accuracy for the test data, which indicates that the extension to multimodal learning alone does not improve the prediction accuracy. For multimodal learning, we need to carefully consider factors such as the task to be solved with the multimodal model, the relationship between the modalities, and the quality of the data, because several previous studies have shown that the prediction accuracy in multimodal learning can be inﬂuenced by such factors \cite{multimodal-notalways}.

Second, in the comparison of four pre-trained models, note that upgrading to the BERT Large model did not always lead to a signiﬁcant improvement in accuracy. Regarding the scale of the model, there is a tendency for an increase in the number of epochs required for model training with bert-large-japanese-v2, and the X1-modality has reached 190 epochs.  This result might indicate the potential for further improvement with an increase in the number of epochs; however,  large model is not always necessary because other models have shown higher prediction accuracy in fewer epochs.

\begin{table*}[t] 
      \begin{center}
      \caption{Result I (Accuracy in Train and Test Data)}
      \label{tab:results1}
        \scalebox{1}{
        \begin{threeparttable}
\begin{tabularx}{\linewidth}{lcccccccccc}
\toprule
 & \multicolumn{3}{c}{Train} & \multicolumn{3}{c}{Test}  & \multicolumn{3}{c}{Epochs} \\
\cmidrule(lr){2-4} \cmidrule(lr){5-7} \cmidrule(lr){8-10} 
BERT Model / Modality &      Both &     X1 &     X2 &     Both &     X1 &     X2&     Both &     X1 &     X2 \\
\midrule
cl-tohoku/bert-base-japanese-v3       &     0.705 &  0.708 &   -     &  {\bf0.711} &  0.703 &  -      &           ~36 &   ~~38~ &   -    \\
cl-tohoku/bert-base-japanese-char-v3  &     0.683 &  0.686 &   -     &     0.690 &  0.678 &    -    &           ~41 &   ~~42~ &  -     \\
cl-tohoku/bert-large-japanese-v2      &  {\bf0.712} &  0.698 &    -    &    0.695 &  0.699 &    -    &           ~95 &  190~ &   -    \\
cl-tohoku/bert-large-japanese-char-v2~~~ &     0.681 &  0.658 &   -     &     0.690 &  0.703 &   -     &           ~58~ &   ~~{\bf18}~ &   -    \\
None                                  &     -      &   -     &  0.555 &      -    &   -     &  0.623 &       -         &     -   &  ~~29~~ \\
\bottomrule
\end{tabularx}
\begin{tablenotes}[para,flushleft,online,normal] 
{\it Note.} Numbers in bold represent the best accuracy in training and testing and the best epochs.
\end{tablenotes}
\end{threeparttable}
        }
        \end{center}
\end{table*} 

\begin{table*}[t] 
      \begin{center}
      \caption{Result II (Group Average)}
      \label{tab:results2}
        \scalebox{1}{
        \begin{threeparttable}
\begin{tabular}{lcccclcccc}
\cmidrule[\heavyrulewidth]{1-4} \cmidrule[\heavyrulewidth]{6-9}
Optimizer & ~~Train~~ & ~~Test~~ & Epochs & ~~~~ & Modality & ~~Train~~ & ~~Test~~ & Epochs \\
\cmidrule{1-4} \cmidrule{6-9}
Adam    &  0.646 & 0.659 & 58.4  & & Both & 0.681 & {\bf0.683} & 70.2\\
Adamax~~~~ &  0.627 & 0.651 & 61.7  &  & X1 & {\bf0.685} & 0.681 & 58.9\\
Nadam & {\bf0.646} & {\bf0.662} & {\bf46.2} & & X2 & 0.546 & 0.604 & {\bf35.5}\\
\cmidrule[\heavyrulewidth]{1-4} \cmidrule[\heavyrulewidth]{6-9}
\end{tabular}
\begin{tablenotes}[para,flushleft,online,normal] 
{\it Note.} Numbers in bold represent the best accuracy in training and testing and the best epochs.
\end{tablenotes}
\end{threeparttable}
        }
        \end{center}
\end{table*} 

In addition, a comparison of the mean accuracies among certain conditions is listed in Table \ref{tab:results2}. In terms of the optimizer, although the training accuracy is almost the same between Adam and Nadam, the latter is better in terms of both test accuracy and the best epoch, which means that, on average, Nadam achieved a higher generalization performance in a shorter time\footnote{Note that the time required for training one epoch did not differ significantly among optimizers.}. In terms of modality, the X1-modal shows high accuracy, as listed in Table 2, for the training process; however, multimodal learning, on average, still shows the highest accuracy for the test data. This result suggests that, on average, multimodal learning improves generalization performance. As expected, the accuracy of the analysis using only X2-modal remains low for both training and testing, although the model converged early.

\section{Conclusions}
This study attempts to construct a multimodal deep learning model that predicts the user ratings of a product using both review text and user profile data simultaneously to account for consumer heterogeneity. First, as academic implications, even when both review and demographic data are relatively small, both the best model and the average score by modality, the prediction accuracy is the best when they are combined, which indicates that multimodal learning that accounts for consumer heterogeneity allows analysis with high robustness and generalizability. Second, it can be shown that, at least when dealing with relatively short sentences such as those used in this study ($len_{max} = 200$), a larger BERT model does not necessarily contribute to an improvement in prediction accuracy. This implies that, particularly in small datasets like those used in this study, converting sentences into word embeddings with BERT is important while the scale of the BERT model is not necessarily critical. Next, the conceptual model presented in this study, as a way to extend review data with cross-sectional data or as a way to extend cross-sectional data with review data, has a potential to be extended to various prediction models in marketing analysis with higher prediction accuracy, compared to conventional methods.

Finally, owing to the constraints of data collection, this study relies on consumer ratings as a proxy for behavioral loyalty and predicts whether it is high or not using the proposed model. However, this methodology can be extended to purchase prediction models by incorporating data that include purchase history and more demographics. 
A few models in previous studies \cite{alaraj} fused modalities twice within a model through the use of attention mechanisms and feature fusion, which aims to enhance prediction accuracy and robustness. Moreover, regarding the actual analysis, further improvements in accuracy can be expected by adopting techniques such as dropout \cite{dropout}. In addition, although the multimodal learning model developed in this study, which utilizes actual review texts as a source of information for understanding consumer heterogeneity, is based on the assumption that consumer heterogeneity is embedded in the review texts, the actual causal relationships need to be carefully examined for the presence of potential endogeneity between the variables.

\section*{Acknowledgements}
All computations in this study were conducted using RAIDEN, which is a computational infrastructure hosted by RIKEN AIP. We would like to express our gratitude to all the members of AIP who maintain the system.

\bibliographystyle{unsrt}
\bibliography{reviewModal}

\begin{thebibliography}{10}

\bibitem{handcraftedfeatures}
Loris Nanni, Stefano Ghidoni, and Sheryl Brahnam.
\newblock Handcrafted vs. non-handcrafted features for computer vision
  classification.
\newblock {\em Pattern Recognition}, 71:158--172, 2017.

\bibitem{featureselection}
Yoshua Bengio.
\newblock Deep learning of representations: Looking forward.
\newblock In {\em International Conference on Statistical Language and Speech
  Processing}, pages 1--37. Springer, 2013.

\bibitem{s2sLSTM-segmentation}
Licheng Zhao, Yi~Zuo, and Katsutoshi Yada.
\newblock Sequential classification of customer behavior based on
  sequence-to-sequence learning with gated-attention neural networks.
\newblock {\em Advances in Data Analysis and Classification}, pages 1--33,
  2022.

\bibitem{dnn-clv}
Rafet Sifa, Julian Runge, Christian Bauckhage, and Daniel Klapper.
\newblock Customer lifetime value prediction in non-contractual freemium
  settings: Chasing high-value users using deep neural networks and smote.
\newblock 2018.

\bibitem{dnn-clv2}
Pei~Pei Chen, Anna Guitart, Ana~Fern{\'a}ndez del R{\'\i}o, and Africa
  Peri{\'a}nez.
\newblock Customer lifetime value in video games using deep learning and
  parametric models.
\newblock In {\em 2018 IEEE international conference on big data (big data)},
  pages 2134--2140. IEEE, 2018.

\bibitem{dnn_purchase}
Jan Valendin, Thomas Reutterer, Michael Platzer, and Klaudius Kalcher.
\newblock Customer base analysis with recurrent neural networks.
\newblock {\em International Journal of Research in Marketing},
  39(4):988--1018, 2022.

\bibitem{dnn_purchase2}
Arthur Toth, Louis Tan, Giuseppe Di~Fabbrizio, and Ankur Datta.
\newblock Predicting shopping behavior with mixture of rnns.
\newblock In {\em eCOM@ SIGIR}, 2017.

\bibitem{dnn_purchase_in_session}
Long Guo, Lifeng Hua, Rongfei Jia, Binqiang Zhao, Xiaobo Wang, and Bin Cui.
\newblock Buying or browsing?: Predicting real-time purchasing intent using
  attention-based deep network with multiple behavior.
\newblock In {\em Proceedings of the 25th ACM SIGKDD international conference
  on knowledge discovery \& data mining}, pages 1984--1992, 2019.

\bibitem{dnn_churn1}
C~Gary Mena, Arno De~Caigny, Kristof Coussement, Koen~W De~Bock, and Stefan
  Lessmann.
\newblock Churn prediction with sequential data and deep neural networks. a
  comparative analysis.
\newblock {\em arXiv preprint arXiv:1909.11114}, 2019.

\bibitem{dnn_churn2}
Philip Spanoudes and Thomson Nguyen.
\newblock Deep learning in customer churn prediction: unsupervised feature
  learning on abstract company independent feature vectors.
\newblock {\em arXiv preprint arXiv:1703.03869}, 2017.

\bibitem{lstm_directMarketing}
Mainak Sarkar and Arnaud De~Bruyn.
\newblock Lstm response models for direct marketing analytics: Replacing
  feature engineering with deep learning.
\newblock {\em Journal of Interactive Marketing}, 53(1):80--95, 2021.

\bibitem{hetero}
Peter~E Rossi, Robert~E McCulloch, and Greg~M Allenby.
\newblock The value of purchase history data in target marketing.
\newblock {\em Marketing Science}, 15(4):321--340, 1996.

\bibitem{hetero3}
Werner~J Reinartz and Vita Kumar.
\newblock The impact of customer relationship characteristics on profitable
  lifetime duration.
\newblock {\em Journal of marketing}, 67(1):77--99, 2003.

\bibitem{multimodal1}
Nitish Srivastava and Russ~R Salakhutdinov.
\newblock Multimodal learning with deep boltzmann machines.
\newblock {\em Advances in neural information processing systems}, 25, 2012.

\bibitem{multimodal2}
Jiquan Ngiam, Aditya Khosla, Mingyu Kim, Juhan Nam, Honglak Lee, and Andrew~Y
  Ng.
\newblock Multimodal deep learning.
\newblock In {\em ICML}, 2011.

\bibitem{niimi_jsai}
Junichiro Niimi and Takahiro Hoshino.
\newblock Predicting purchases with using the variety of customer behaviors
  -analysis of the purchase history and the browsing history by deep learning-.
\newblock {\em Transactions of the Japanese Society for Artificial
  Intelligence}, 32(2):B--G63\_1--9, 2017.

\bibitem{bert}
Jacob Devlin, Ming-Wei Chang, Kenton Lee, and Kristina Toutanova.
\newblock Bert: Pre-training of deep bidirectional transformers for language
  understanding.
\newblock {\em arXiv preprint arXiv:1810.04805}, 2018.

\bibitem{distilbert}
Victor Sanh, Lysandre Debut, Julien Chaumond, and Thomas Wolf.
\newblock {DistilBERT, a distilled version of BERT: smaller, faster, cheaper
  and lighter}.
\newblock {\em arXiv}, 2019.

\bibitem{roberta}
Yinhan Liu, Myle Ott, Naman Goyal, Jingfei Du, Mandar Joshi, Danqi Chen, Omer
  Levy, Mike Lewis, Luke Zettlemoyer, and Veselin Stoyanov.
\newblock Roberta: A robustly optimized bert pretraining approach.
\newblock {\em arXiv preprint arXiv:1907.11692}, 2019.

\bibitem{openCalm}
Alex Andonian, Quentin Anthony, Stella Biderman, Sid Black, Preetham Gali, Leo
  Gao, Eric Hallahan, Josh Levy-Kramer, Connor Leahy, Lucas Nestler, Kip
  Parker, Michael Pieler, Shivanshu Purohit, Tri Songz, Wang Phil, and Samuel
  Weinbach.
\newblock {GPT-NeoX: Large Scale Autoregressive Language Modeling in PyTorch},
  8 2021.

\bibitem{clTohoku}
Tohoku~NLP Group.
\newblock cl-tohoku/bert-japanese (github:
  https://github.com/cl-tohoku/bert-japanese), 2023.

\bibitem{word2vec}
Tomas Mikolov, Kai Chen, Greg Corrado, and Jeffrey Dean.
\newblock Efficient estimation of word representations in vector space.
\newblock {\em arXiv preprint arXiv:1301.3781}, 2013.

\bibitem{doc2vec}
Quoc Le and Tomas Mikolov.
\newblock Distributed representations of sentences and documents.
\newblock In {\em International conference on machine learning}, pages
  1188--1196. PMLR, 2014.

\bibitem{GooglePlay}
Zeynep~Hilal Kilimci.
\newblock Prediction of user loyalty in mobile applications using deep
  contextualized word representations.
\newblock {\em Journal of Information and Telecommunication}, 6(1):43--62,
  2022.

\bibitem{RNN}
David~E Rumelhart, Geoffrey~E Hinton, and Ronald~J Williams.
\newblock Learning representations by back-propagating errors.
\newblock {\em nature}, 323(6088):533--536, 1986.

\bibitem{CNN}
Yann LeCun, L{\'e}on Bottou, Yoshua Bengio, and Patrick Haffner.
\newblock Gradient-based learning applied to document recognition.
\newblock {\em Proceedings of the IEEE}, 86(11):2278--2324, 1998.

\bibitem{LSTM}
Sepp Hochreiter and J{\"u}rgen Schmidhuber.
\newblock Long short-term memory.
\newblock {\em Neural computation}, 9(8):1735--1780, 1997.

\bibitem{bert_hotel}
Yuanyuan Zhuang and Jaekyeong Kim.
\newblock A bert-based multi-criteria recommender system for hotel promotion
  management.
\newblock {\em Sustainability}, 13(14):8039, 2021.

\bibitem{bert_influencer}
Seungbae Kim, Xiusi Chen, Jyun-Yu Jiang, Jinyoung Han, and Wei Wang.
\newblock Evaluating audience loyalty and authenticity in influencer marketing
  via multi-task multi-relational learning.
\newblock In {\em Proceedings of the International AAAI Conference on Web and
  Social Media}, volume~15, pages 278--289, 2021.

\bibitem{multimodal-survey}
Dhanesh Ramachandram and Graham~W Taylor.
\newblock Deep multimodal learning: A survey on recent advances and trends.
\newblock {\em IEEE signal processing magazine}, 34(6):96--108, 2017.

\bibitem{alaraj}
Maher Ala'raj, Maysam~F Abbod, and Munir Majdalawieh.
\newblock Modelling customers credit card behaviour using bidirectional lstm
  neural networks.
\newblock {\em Journal of Big Data}, 8(1):1--27, 2021.

\bibitem{hetero1}
Carsten Hahn, Michael~D Johnson, Andreas Herrmann, and Frank Huber.
\newblock Capturing customer heterogeneity using a finite mixture pls approach.
\newblock {\em Schmalenbach Business Review}, 54:243--269, 2002.

\bibitem{hetero2}
Thomas Otter, Regina T{\"u}chler, and Sylvia Fr{\"u}hwirth-Schnatter.
\newblock Capturing consumer heterogeneity in metric conjoint analysis using
  bayesian mixture models.
\newblock {\em International Journal of Research in Marketing}, 21(3):285--297,
  2004.

\bibitem{review_for_heterogeneity}
Silvana Aciar, Debbie Zhang, Simeon Simoff, and John Debenham.
\newblock Recommender system based on consumer product reviews.
\newblock In {\em 2006 IEEE/WIC/ACM International Conference on Web
  Intelligence (WI 2006 Main Conference Proceedings)(WI'06)}, pages 719--723.
  IEEE, 2006.

\bibitem{adam}
Diederik~P Kingma and Jimmy Ba.
\newblock Adam: A method for stochastic optimization.
\newblock {\em arXiv preprint arXiv:1412.6980}, 2014.

\bibitem{nadam}
Timothy Dozat.
\newblock Incorporating nesterov momentum into adam.
\newblock 2016.

\bibitem{earlyStopping}
Lutz Prechelt.
\newblock Early stopping-but when?
\newblock In {\em Neural Networks: Tricks of the trade}, pages 55--69.
  Springer, 1998.

\bibitem{multimodal-notalways}
Douwe Kiela and L{\'e}on Bottou.
\newblock Learning image embeddings using convolutional neural networks for
  improved multi-modal semantics.
\newblock In {\em Proceedings of the 2014 Conference on empirical methods in
  natural language processing (EMNLP)}, pages 36--45, 2014.

\bibitem{dropout}
Nitish Srivastava, Geoffrey Hinton, Alex Krizhevsky, Ilya Sutskever, and Ruslan
  Salakhutdinov.
\newblock Dropout: a simple way to prevent neural networks from overfitting.
\newblock {\em The journal of machine learning research}, 15(1):1929--1958,
  2014.

\end{thebibliography}

\end{document}